\documentclass{article}
\usepackage[utf8]{inputenc}

\usepackage[colorlinks=true,linkcolor=black,anchorcolor=blue,citecolor=blue,filecolor=blue,menucolor=blue,runcolor=blue,urlcolor=blue]{hyperref}
\usepackage{xcolor}
\usepackage{amssymb,amsmath,amsfonts}
\usepackage{cancel}
\usepackage{comment}
\usepackage{epsfig}
\usepackage{graphicx}
\usepackage{epstopdf}
\usepackage{caption}
\usepackage{subcaption}
\usepackage{amscd}
\usepackage{amsthm}
\usepackage{latexsym}
\usepackage{cite}
\usepackage{amsbsy}
\usepackage{bbm}
\usepackage[english]{babel}
\usepackage{psfrag}
\usepackage{tabularx}
\usepackage{enumerate}
\usepackage[normalem]{ulem}

\allowdisplaybreaks[1]

\usepackage{overpic}
\usepackage{array}
\usepackage{rotating}
\usepackage{lipsum}
\usepackage{color}

\newcommand{\be}{\begin{equation}}
\newcommand{\ee}{\end{equation}}
\newcommand{\bea}{\begin{eqnarray}}
\newcommand{\eea}{\end{eqnarray}}

\newcommand{\Vierh}[2]{\hat{E}_{#1}{}^{#2}}
\newcommand{\Viel}[3]{{#1}_{#2}{}^{#3}}

\newcommand{\CA}{{\cal A}}
\newcommand{\CB}{{\cal B}}
\newcommand{\CF}{{\cal F}}

\newcommand{\lr}{\left (}
\newcommand{\rr}{\right )}
\newcommand{\ls}{\left [}
\newcommand{\rs}{\right ]}

\newcommand{\p}{\partial}
\newcommand{\NC}{Newton-Cartan }

\title{An Action for Extended String Newton-Cartan Gravity}
\author{Eric A.~Bergshoeff${}^{\, a}$, Kevin T.~Grosvenor${}^{\, b}$, Ceyda \c{S}im\c{s}ek${}^{\, a}$, Ziqi Yan${}^{\, c}$\\[.5truecm]
${}^a$Van Swinderen Institute, University of Groningen\\
Nijenborgh 4, 9747 AG Groningen, The Netherlands\medskip\\
${}^{\, b}$Institut f\"ur Theoretische Physik und Astrophysik,\\
Julius-Maximilians-Universit\"at W\"urzburg, \\ Am Hubland, 97074 Würzburg, Germany\medskip\\
${}^c$Perimeter Institute for Theoretical Physics\\
31 Caroline St N, Waterloo, ON N2L 6B9, Canada\\[.5truecm]
{\sl E-mail:} E.A.Bergshoeff@rug.nl, kevinqg1@gmail.com,\\   c.simsek@rug.nl, zyan@pitp.ca}
\date{October 22 2018}

\begin{document}

\maketitle

\centerline{ABSTRACT}

We construct an action for four-dimensional \emph{extended string \NC gravity} which is an extension of the string Newton-Cartan gravity that underlies nonrelativistic string theory. The action can be obtained as a nonrelativistic limit of the Einstein-Hilbert action in General Relativity augmented with a term that contains an auxiliary two-form and one-form gauge field that both have zero flux on-shell. The four-dimensional extended string \NC gravity is based on a central extension of the algebra that underlies string \NC gravity.

The construction is  similar to the earlier construction of a three-dimensional Chern-Simons  action for extended \NC gravity, which is based on a central extension of the algebra that underlies Newton-Cartan gravity. We show that this three-dimensional action is naturally obtained from the four-dimensional action by  a reduction  over the spatial isometry direction longitudinal to the string followed by a truncation of the extended string \NC gravity fields.  Our construction can be seen as a special case of the construction  of an action  for {\sl extended $p$-brane \NC gravity} in $p+3$ dimensions.
\newpage

\tableofcontents

\section{Introduction}

 String Newton-Cartan (NC) gravity is an extension of NC gravity. In NC gravity, there is a one-dimensional foliation direction of NC spacetime corresponding to the absolute time direction longitudinal to the worldline of a particle. In string NC gravity, this one-dimensional foliation structure is replaced by a two-dimensional foliation with the two (timelike and spatial) foliation directions longitudinal to the world-sheet of a string. String NC gravity is the nonrelativistic gravity that naturally arises in the context of nonrelativistic string theory \cite{Gomis:2000bd,Bergshoeff:2018yvt}.\,\footnote{In this paper we  consider a nonrelativistic string theory that is nonrelativistic in the target space but relativistic on the worldsheet. For other recent work on nonrelativistic strings, see \cite{Batlle:2016iel,Gomis:2016zur,Batlle:2017cfa,Harmark:2017rpg,Kluson:2018egd,Harmark:2018cdl}. In \cite{Harmark:2017rpg,Kluson:2018egd,Harmark:2018cdl}, for zero torsion,  a specific truncation of the string NC gravity in the target space was considered, which leads to NC gravity in one dimension lower, supplemented with an extra worldsheet scalar parametrizing the spatial foliation direction.} It also arises in  the study of nonrelativistic holography with nonrelativistic  gravity in the bulk where it could be used to probe nonrelativistic  conformal field theories (CFTs) at the boundary with (an infinite extension of) Galilean conformal symmetries \cite{Bagchi:2009my}.

The transformation rules and equations of motion of string NC gravity in any dimension can be obtained by taking a nonrelativistic limit of the transformation rules and Einstein equations of General Relativity augmented with a zero-flux two-form gauge field $\hat{\CB}_{\mu\nu}$ \cite{Gomis:2000bd,inpreparation}.\,\footnote{We denote relativistic fields with a hat to distinguish them from the nonrelativistic fields that will be introduced later.} This same two-form gauge field couples to the string via a Wess-Zumino (WZ) term when taking the nonrelativistic limit of string theory. This WZ term is crucial to cancel infinities that otherwise would arise when taking the nonrelativistic limit of string theory \cite{Gomis:2000bd}.\,\footnote{Ignoring boundary conditions this WZ term is a total derivative. Including boundary conditions the WZ  term is essential to obtain string winding modes in the spectrum of the nonrelativistic string \cite{Gomis:2000bd}. Independent of this, the two-form gauge field is also crucial to derive the off-shell transformation rules of string NC gravity as a limit of those of General Relativity \cite{inpreparation}.} Taking the nonrelativistic  limit of the off-shell Einstein-Hilbert term, one again obtains divergent terms.  Unlike the on-shell case,  these divergent terms cannot be cancelled by making use of a two-form gauge-field alone. Using a first-order formulation, we will show in this work that, in four-dimensional spacetime, the divergent terms that arise from the Einstein-Hilbert term can be cancelled by  adding the following BF term to the Einstein-Hilbert term  that contains not only  the two-form gauge field $\hat{\CB}_{\mu\nu}$ but also  an additional one-form gauge field $\hat{\CA}_{\mu}$:
\be \label{eq:relaction}
    \hat{S} = \frac{1}{2\kappa^2} \int d^4 x \lr \hat{E} \, \hat{R} -  \epsilon^{\mu\nu\rho\sigma} \hat{\CB}_{\mu\nu} \partial_\rho \hat{\CA}_{\sigma} \rr \,.
\ee
 Here $\kappa$ is the gravitational coupling constant, $\hat{R}$ is the Ricci scalar  and $\hat{E}$ is the determinant of the Vierbein field $\Vierh{\mu}{\hat{A}}$. Both gauge fields have zero flux on-shell. They transform under gauge transformations with parameters   $\hat{\eta}_\mu$ and $\hat{\zeta}$ as follows:
\be \label{eq:gtrnsf}
    \delta \hat{\CB}_{\mu\nu} = \p_\mu \hat{\eta}_\nu - \p_\nu \hat{\eta}_\mu\,,
        \qquad
    \delta \hat{\CA}_\mu = \p_\mu \hat{\zeta}\,.
\ee
The Vierbein and the  independent spin-connection $\Omega$ transform under Lorentz transformations in the usual way
\be \label{eq:gtrnsf2}
\delta \hat{E}_\mu{}^{\hat{A}} = \hat{\Lambda}^{\hat{A}}{}_{\hat{B}}\hat{E}_\mu{}^{\hat{B}}\,,\hskip 1truecm
\delta \hat{\Omega}_\mu{}^{\hat{A}\hat{B}} = \partial_\mu\hat{\Lambda}^{\hat{A}\hat{B}}  -2 \hat{\Omega}_\mu{}^{[\hat{A}}{}_{\hat{C}} \Lambda^{\hat{C}\hat{B}]}\,.
\ee
All fields transform as tensors under general coordinate transformations with parameters $\hat{\Xi}^\mu$.

Due to the presence of the additional gauge field $\hat{\CA}_{\mu}$ needed to write down the second term in the action eq.~\eqref{eq:relaction} one obtains, after taking the limit, not an action for string NC gravity but for a different nonrelativistic gravity theory which we call Extended String \NC (ESNC) gravity. This ESNC gravity  theory is based on a central extension of the algebra that underlies string NC gravity.  We will call this extended algebra the ESNC algebra.

The above construction of a four-dimensional action for ESNC gravity  is similar to an earlier construction of three-dimensional extended NC gravity\,\footnote{This three-dimensional extended NC gravity is  referred to as the ``extended Bargmann gravity" in \cite{Bergshoeff:2016lwr}.} \cite{Papageorgiou:2009zc,Bergshoeff:2016lwr,Hartong:2016yrf}. The main difference is that in the three-dimensional  case one defines a limit leading to a one-dimensional foliation and the two-form gauge field gets replaced by a one-form gauge field $\hat{\CB}_{\mu}$ leading to the following three-dimensional analogue of the four-dimensional action  eq.~\eqref{eq:relaction}:
\be \label{secondterm3D}
    \hat{S} = \frac{1}{2\kappa^2} \int d^3 x \lr \hat{E} \, \hat{R} -  \epsilon^{\mu\nu\rho} \hat{\CB}_{\mu} \partial_\nu \hat{\CA}_{\rho} \rr \,.
\ee
The common feature between the 3D and 4D actions is that the extra term we add to the Einstein-Hilbert term contains a $p$-form $\hat{\CB}_{\mu_1\cdots \mu_{p+1}}$ that naturally couples to a $p$-brane and a vector $\hat{\CA}_\mu$ representing a central extension of the underlying  algebra. Clearly, such a term can only be written down in $p+3$ dimensions which is precisely the dimension in which a $p$-brane has {\sl two} transverse directions.

In this work we will derive the transformation rules of the ESNC gravity fields and their curvatures  by gauging the ESNC algebra. A standard part of this gauging procedure is to impose a set of so-called conventional curvature constraints  expressing some of the gauge fields in terms of a set of independent fields. Usually, one imposes a maximal set of constraints in order to make the resulting theory irreducible. We will not do so here since we will take the nonrelativistic limit of first-order general relativity leading to a first-order nonrelativistic action whose equations of motion will automatically lead to a set of curvature constraints. It turns out  that these curvature constraints are not identical to the maximal set of conventional constraints one could impose. As a consequence of this, we will find that not all components of the boost spin-connection fields can be solved for.

It is instructive to contrast  this situation with first-order general relativity described by the action
\be \label{eq:EH}
    \hat{S} = \frac{1}{2\kappa^2} \int d^4 x \, \hat{E} \, \hat{R} (\hat{\Omega})\,.
\ee
In that case the equations of motion of the independent spin-connection fields  yield precisely the maximum set of 24  constraints on the curvatures corresponding to the spacetime generators $\hat P$  that one imposes in the gauging procedure:
\be
\hat{R}_{\mu\nu}{}^{\hat{A}}(\hat{P}) = 0\,.
\ee
These 24 curvature constraints can be used to solve for all 24 components of the spin-connection fields $\hat{\Omega}_\mu{}^{\hat{A}\hat{B}}$.  In the relativistic case, there are no geometrical constraints, i.e.~constraints on the independent timelike Vierbein fields. In contrast, we find that in the nonrelativistic case the components of the boost spin-connection fields that remain independent occur as Lagrange multipliers in the nonrelativistic action  imposing some (but not all) of the geometric constraints of ESNC gravity.\,\footnote{The remaining geometric constraints are obtained by varying the central extension gauge field, see the second-order nonrelativistic action eq.~\eqref{second-order}.} This is very similar to what happens in the case of the so-called Carroll and Galilei gravity theories \cite{Bergshoeff:2017btm}.

Our work should be contrasted with the Post-Newtonian (PN)  approximation to general relativity where one expands the Vielbein in powers of the speed of light $c$  without introducing additional gauge fields like we do in this work. In particular, it has been recently shown that a large $c$ expansion of general relativity coupled to matter can give rise to torsional NC gravity  \cite{VandenBleeken:2017rij} and nonrelativistic actions \cite{Hansen:2018ofj}. The difference with this work is that  in the PN approximation one  considers  truncations   and  general relativity coupled to (uncharged) matter whereas in this work we take a limit  that is  appropriate when considering  sourceless (ignoring back-reaction) Einstein equations or string actions with the string charged under a two-form gauge field.

The organization of this work is as follows.
In section \ref{sec:gauging} we will derive the transformation rules and curvatures of the independent gauge fields  by gauging the  4D  ESNC algebra.
These ingredients will be needed to  describe the 4D ESNC gravity theory.
In section \ref{sec:action}, we will consider the relativistic action eq.~\eqref{eq:relaction} containing the extra two-form and one-form gauge fields. First, we will show how an action  for the 4D ESNC gravity theory can be obtained by taking  the  nonrelativistic limit of the action  eq.~\eqref{eq:relaction}.
Next, we will show how  the nonrelativistic rules we derived in the previous section can be obtained by taking the limit of the relativistic transformations rules.
   In the same section we will  obtain the second-order formulation of the 4D ESNC gravity theory and present the  equations of motion. In section \ref{sec:3D}, we will show how, by truncation, the ESNC algebra is related to the 3D algebra that underlies extended NC gravity. Furthermore, we will show how the action describing 4D ESNC gravity  reduces to the 3D nonrelativistic action for extended NC gravity given in  \cite{Bergshoeff:2016lwr,Hartong:2016yrf}. We will also briefly discuss the possibility of further extending the ESNC algebra.
 Finally, in section \ref{sec:concl} we will discuss several issues that follow from our results. In appendix \ref{app:irreducible} we will present the maximal set of conventional constraints that one can impose when gauging the ESNC algebra without requiring an action. In appendix \ref{app:pbrane} we will discuss the generalization of ESNC to an extended $p$-brane NC algebra in $p+3$ dimensions. In appendix \ref{app:form} we will show that, although there is an action principle for the ESNC gravity theory, the corresponding ESNC algebra does not allow a  nondegenerate symmetric invariant bilinear form.

\section{Gauging the 4D Extended String \NC Algebra} \label{sec:gauging}

We start by writing down the symmetries and generators of the 4D extended string \NC algebra. For this purpose  we split the flat index $\hat A$ of General Relativity into an index $A\,  (A=0,1)$ longitudinal to the string and an index $A^\prime \, (A' = 2,3)$ transverse to the string.

In \cite{Brugues:2004an,Andringa:2012uz} it has been shown that the underlying algebra of string NC gravity is an extension of the string Galilei algebra that includes the non-central extensions $Z_A$ and $Z$\,. In this paper, we will refer to this extended string Galilei algebra as the \emph{string Newton-Cartan} algebra, which conveniently implies that this is the algebra that underlies string NC gravity. It turns out that the 4D string NC algebra allows an additional  central extension whose generator we denote by $S$. This central extension is the analogue of the second central extension of the 3D Bargmann algebra
\cite{book}.\,\footnote{The physics behind this central extension, in particular its relation to (anyon) spin, has been discussed in \cite{Duval:2000xr,Jackiw:2000tz}.}\,\footnote{Following the logic of referring to the  algebra underlying  string NC gravity as the string NC algebra, we refer to the Bargmann algebra that underlies NC gravity as the  the ``Newton-Cartan algebra." See section \ref{sec:3D}.} We thus end up with the following symmetries and corresponding generators:
\begin{align*}
    \text{longitudinal translations} \qquad & H_A \\
    \text{transverse translations} \qquad & P_{A'} \\
    \text{longitudinal Lorentz transformations} \qquad & M \\
    \text{string Galilei boosts} \qquad & G_{AA'} \\
    \text{spatial rotations} \qquad & J \\
    \text{non-central extensions} \qquad & Z_A\ \textrm{and} \ Z \\
    \text{central extension} \qquad & S
\end{align*}
We use the following terminology for the algebras that are formed by different sets of the above generators:
\begin{enumerate}

\item

	\emph{String Galilei algebra} consists of the generators $H_A\,, P_{A'}\,, M\,, G_{AA'}$ and $J$\,.
	
\item

	\emph{String Newton-Cartan algebra} is a noncentral extension of string Galilei algebra that includes the generators $Z_A$ and $Z$\,. This algebra underlies string NC gravity.
	
\item

	\emph{Extended String Newton-Cartan (ESNC) algebra} is a central extension of the 4D string NC algebra in four dimensions that includes the generator $S$\,. This algebra underlies ESNC gravity.

\end{enumerate}
We emphasize that the string Galilei algebra and the string NC algebra exist in any dimension, but that the ESNC algebra only exists in four dimensions.

The non-zero commutators among the generators of the ESNC algebra are given by\footnote{In appendix \ref{app:pbrane} we will show that there is a natural $p$-brane generalization of the ESNC algebra in $p+3$ dimensions.}
\begin{subequations} \label{eq:ESNCa}
\begin{align}
    [H_A, M] & = \epsilon_A{}^B H_B\,, \hskip 2truecm
    [H_A, G_{BA'}]  = \eta^{}_{AB} P_{A'}\,, \\
    [P_{A'}, J] & = \epsilon_{A'}{}^{B'} P_{B'}\,, \hskip 2truecm
    [G_{AA'}, M] = \epsilon_A{}^B G_{BA'}\,, \\
    [G_{AA'}, J] & = \epsilon_{A'}{}^{B'} G_{AB'}\,,
\intertext{and}
    [G_{AA'}, P_{B'}] & = \delta_{A'\!B'} Z_A\,, \label{eq:ESNCaGP}\\
    [G_{AA'}, G_{BB'}] & = \delta_{A'\!B'} \epsilon_{AB} Z + \epsilon_{A'\!B'} \eta_{AB} S\,; \label{eq:ESNCaGG}\\
    [Z_A, M] & = \epsilon_A{}^B Z_B\,, \hskip 2truecm
    [H_A, Z]  = \epsilon_A{}^B Z_B\,.
\end{align}
\end{subequations}
We have taken the following convention for the Levi-Civita symbols:
\be
	\epsilon_{01} = - \epsilon_{10} = \epsilon_{23} = - \epsilon_{32} = 1\,.
\ee
The index $A$ can be raised by using the Minkowskian metric
\be
	\eta^{AB} =
		\begin{pmatrix}
			-1 & \,\,0 \\
			0 & \,\,1
		\end{pmatrix}\,,
\ee
and the index $A'$ can be raised by the Kronecker symbol $\delta^{A'B'}$\,.

We introduce a Lie algebra valued gauge field $\Theta_\mu$ that associates to each of the generators of the ESNC algebra a corresponding gauge field as follows:
\be
	\Theta_\mu = H_A \tau_\mu{}^A + P_{A'} E_\mu{}^{A'} + G_{AA'} \Omega_\mu{}^{AA'} + M \Sigma_\mu + J \Omega_\mu + Z_A m_\mu{}^A + Z n_\mu + S s_\mu\,.
\ee
Note that $\Sigma_\mu$ is the longitudinal spin-connection while $\Omega_\mu$ is the transverse spin-connection. Considering the symmetries, we ignore the (longitudinal and transverse) translations and, instead,  declare that all gauge fields transform as covariant vectors under reparametrizations with parameter $\xi^\mu(x)$.\,\footnote{The reason for doing this is that in the next section we will  take the nonrelativistic limit of  general relativity in a first-order formulation.  The 4D Einstein-Hilbert action in such a first-order formulation is invariant under Lorentz transformations and general coordinate transformations only \cite{Banados:1996hi}.}
Under the remaining gauge transformations
\begin{equation}
\delta \Theta_\mu=\partial_\mu \Lambda - [\Theta_\mu,\Lambda]\,,
\end{equation}
with
\begin{equation}
\Lambda = G_{AA'} \lambda^{AA'} + M \lambda + J \lambda' + Z_A \sigma^A + Z \sigma + S \rho\,,
\end{equation}
the (independent) gauge fields transform as follows:
\begin{subequations} \label{eq:trnsfs}
\begin{align}
	\delta \tau_\mu{}^A & = \lambda \, \epsilon^A{}_B \tau_\mu{}^B \,, \qquad\qquad
	\delta E_\mu{}^{A'} = - \lambda_A{}^{A'} \tau_\mu{}^A + \lambda' \epsilon^{A'}{}_{B'}  E_\mu{}^{B'} \,, \\[4pt]
	\delta \Sigma_\mu & = \p_\mu \lambda\,, \qquad\qquad\qquad\quad\,\,\,
	\delta \Omega_\mu = \p_\mu \lambda'\,, \\[4pt]
	\delta \Omega_\mu{}^{AA'} & = \p_\mu \lambda^{AA'} - \epsilon^A{}_B \lambda^{BA'} \Sigma_\mu - \epsilon^{A'}{}_{B'} \lambda^{AB'} \Omega_\mu + \lambda \, \epsilon^A{}_B \Omega_\mu{}^{BA'} + \lambda' \, \epsilon^{A'}{}_{B'} \Omega_\mu{}^{AB'} \,, \\[4pt]
	\delta n_\mu & = \p_\mu \sigma + \epsilon_{AB} \lambda^{AA'} \Omega_\mu{}^{BA'}\,, \\[4pt]
	\delta s_\mu & = \p_\mu \rho + \epsilon_{A'B'} \lambda^{AA'} \Omega_{\mu A}{}^{B'}\,, \\[4pt]
	\delta m_\mu{}^A & = \p_\mu \sigma^A - \epsilon^A{}_B \sigma^B \Sigma_\mu + \lambda \, \epsilon^A{}_B m_\mu{}^B + \lambda^{AA'} E_{\mu A'} + \epsilon^A{}_B \tau_\mu{}^B \sigma\,.
\end{align}
\end{subequations}

The curvature two-form $\CF_{\mu\nu}$ associated with $\Theta_\mu$ is
\begin{align}
	\CF_{\mu\nu} & = \p_\mu \Theta_\nu - \p_\nu \Theta_\mu - [\Theta_\mu, \Theta_\nu] \notag \\
		& = H_A R_{\mu\nu}{}^A (H) + P_{A'} R_{\mu\nu}{}^{A'} (P) + G_{AA'} R_{\mu\nu}{}^{AA'} (G) + M R_{\mu\nu} (M) + J R_{\mu\nu} (J) \notag \\
		& \quad + Z_A R_{\mu\nu}{}^A(Z) + Z R_{\mu\nu} (Z) + S R_{\mu\nu} (S)\,,
\end{align}
with the expressions for the curvature two-forms given by
\begin{subequations}
\begin{align}
	R_{\mu\nu}{}^A (H) & = 2 \lr \p^{}_{[\mu} \tau^{}_{\nu]}{}^{A} + \epsilon^A{}_{B} \tau^{}_{[\mu}{}^B \Sigma^{}_{\nu]} \rr \,, \\[4pt]
	R_{\mu\nu}{}^{A'} (P) & = 2 \lr \p^{}_{[\mu} E^{}_{\nu]}{}^{A'} + \epsilon^{A'}{}_{B'} E^{}_{[\mu}{}^{B'} \Omega^{}_{\nu]} - \tau^{}_{[\mu}{}^A \Omega^{}_{\nu]A}{}^{A'} \rr\,, \\[4pt]
	R_{\mu\nu}{}^A (Z) & = 2 \lr \p^{}_{[\mu} m^{}_{\nu]}{}^A + \epsilon^A{}_B m^{}_{[\mu}{}^B \Sigma^{}_{\nu]} + \epsilon^A{}_B \tau^{}_{[\mu}{}^B n^{}_{\nu]} + E^{}_{[\mu}{}^{A'} \Omega^{}_{\nu]}{}^A{}_{A'}  \rr \\[4pt]
	R_{\mu\nu} (M) & = 2 \p^{}_{[\mu} \Sigma^{}_{\nu]}\,, \\[4pt]
	R_{\mu\nu} (J) & = 2 \p^{}_{[\mu} \Omega^{}_{\nu]}\,, \\[4pt]
	R_{\mu\nu}{}^{AA'} (G) & = 2 \lr \p^{}_{[\mu} \Omega^{}_{\nu]}{}^{AA'} + \epsilon^A{}_B \Omega_{[\mu}{}^{BA'} \Sigma_{\nu]} + \epsilon^{A'}{}_{B'} \Omega_{[\mu}{}^{AB'} \Omega_{\nu]} \rr \,, \\[4pt]
	R_{\mu\nu} (Z) & = 2 \p^{}_{[\mu} n^{}_{\nu]} - \epsilon_{AB} \Omega^{}_{[\mu}{}^{AA'} \Omega^{}_{\nu]}{}^B{}^{}_{A'} \,, \\[4pt]
	R_{\mu\nu} (S) & = 2 \p^{}_{[\mu} s^{}_{\nu]} - \epsilon_{A'\!B'} \Omega^{}_{[\mu}{}^{AA'} \Omega^{}_{\nu] A}{}^{B'} \,.	
\end{align}
\end{subequations}

At this stage one usually imposes a maximal set of so-called conventional curvature constraints in order to obtain an irreducible gauge theory.
More precisely, one sets all curvatures equal to zero that contain terms of the form gauge field times a  timelike Vierbein $\tau_\mu{}^A$
or a spacelike Vierbein $E_\mu{}^{A^\prime}$. Since these Vierbeine are invertable\,\footnote{See eq.~\eqref{inverses} in the next section for the definition of the inverses.}  this allows one to solve for (some of the components of) the gauge fields that multiply these Vierbeine in terms of the other terms that are contained in the expression of the curvature. However, we will find that  imposing such a maximal set of constraints is not consistent with the existence of an action. For this reason, we will first construct in the next section a first-order nonrelativistic action and next impose only those curvature constraints that follow from the variation of this action. For completeness and to compare, we will give the details of the irreducible gauging (without an action) in appendix \ref{app:irreducible}.

\section{The Action} \label{sec:action}

Our starting point is the first-order action eq.~\eqref{eq:relaction} containing the sum of the Einstein-Hilbert term and a term containing a two-form and one-form gauge field. Before taking the nonrelativistic limit of this action, we first wish to define an expansion of the relativistic fields occurring in the action eq.~\eqref{eq:relaction} in terms of the nonrelativistic ESNC fields defined in the previous section  such that the correct transformation rules are reproduced. This leads us to define the following expansion in terms of a parameter $\omega$ which later will be taken to infinity:
\begin{subequations}\label{expansion}
\begin{align}
    \Vierh{\mu}{A} & = \omega \Viel{\tau}{\mu}{A} + \frac{1}{\omega} \Viel{m}{\mu}{A} \,,
    	&
    \Vierh{\mu}{A'} &= \Viel{E}{\mu}{A'} \,, \\[4pt]
    \hat{\Sigma}_{\mu} &= \Sigma_{\mu}+ \frac{1}{\omega^2} n_\mu \,,
    	&
    \hat{\Omega}_{\mu}{}^{AA'} &= \frac{1}{\omega} \Omega_{\mu}{}^{AA'} \,,
    	&
    \hat{\Omega}_{\mu}&= \Omega_{\mu}- \frac{1}{\omega^2} s_\mu \,, \\[4pt]
     \hat{\CA}_\mu & = \Omega_\mu \,,
	&
    \hat{\CB}_{\mu\nu} & = \omega^2\,\epsilon_{AB} \Viel{\tau}{\mu}{A} \Viel{\tau}{\nu}{B} \,,
  \end{align}
\end{subequations}
where we have written $\hat{A}= (A,A')$ with $ (A=0,1; A'=2,3)$ and
\begin{equation}
 \hat{\Omega}_{\mu}{}^{AB} = \epsilon^{AB}\hat{\Sigma}_\mu\,,\hskip 1.5truecm
\hat{\Omega}_{\mu}{}^{A'B'} = \epsilon^{A'B'}\hat{\Omega}_\mu\,.
\end{equation}
Finally, we expand the relativistic symmetry parameters as follows:\,\footnote{Strictly speaking, we should also expand the g.c.t.~parameters:
\begin{equation}
\hat{\Xi}^{\mu} = \xi^\mu +   \frac{1}{\omega^2}\sigma^\mu\,.
\end{equation}
We find that under reparametrizations with parameter $\xi^\mu$ all nonrelativistic fields transform as covariant vectors and that the next-order parameter $\sigma^\mu$ does not contribute.}
\begin{subequations}\label{expparameters}
\begin{align}
\hat \Lambda &= \lambda - \frac{1}{\omega^2}\sigma\,,\hskip 1.truecm
\hat {\Lambda}^{AA'} = \frac{1}{\omega}\lambda^{AA'}\,,  \hskip 1.1truecm
\hat {\Lambda}^\prime = \lambda^\prime + \frac{1}{\omega^2}\rho\,,\\
\hat {\eta}_\mu &= \epsilon_{AB}\tau_\mu{}^A \sigma^B\,,\hskip 1.5truecm
\hat {\zeta} = \lambda^\prime\,,
\end{align}
\end{subequations}
where we have written
\begin{equation}
\hat {\Lambda}^{AB} = \epsilon^{AB}\hat{\Lambda}\hskip 1truecm \textrm{and} \hskip 1truecm  \hat {\Lambda}^{A' B'} = \epsilon^{A' B'} \hat{\Lambda}^\prime\,.
\end{equation}

First, we substitute the expansion eq.~\eqref{expansion}
into the action eq.~\eqref{eq:relaction}.  To simplify the calculation, it is convenient to work with the following  formulation of the action that avoids the explicit appearance  of inverse Vierbein fields:\,\footnote{The inverse Vierbein fields will be needed later  in the second-order formulation when we  solve for the connection fields.}${}^,$\footnote{The first term can be viewed as a constrained BF term that is central in the Plebanski formulation of general relativity \cite{Plebanski:1977zz}, see, e.g., \cite{Gielen:2010cu}.}
\be \label{eq:relaction2}
    \hat{S} = \frac{1}{2\kappa^2} \int d^4 x \lr
    \frac{1}{4}\epsilon^{\mu\nu\rho\sigma}\epsilon_{\hat{A}\hat{B}\hat{C}\hat{D}} E_{\mu}{}^{\hat{A}}E_{\nu}{}^{\hat{B}} \hat{R}_{\rho\sigma}^{\hspace{0.33cm} \hat{C} \hat{D}}(\hat{\Omega})
        -  \epsilon^{\mu\nu\rho\sigma} \hat{\CB}_{\mu\nu} \partial_\rho \hat{\CA}_{\sigma} \rr \,.
\ee
For $\epsilon^{\mu\nu\rho\sigma}$ we choose the convention $\epsilon^{0123} = 1$ and for $\epsilon_{\hat{A}\hat{B}\hat{C}\hat{D}}$ we choose the convention $\epsilon_{0123} = 1$\,.
After a  straightforward calculation, we find that all divergent terms proportional to $\omega^2$ cancel and that the remaining expression, after taking the limit $\omega\rightarrow \infty$, is given by the following nonrelativistic action describing the ESNC gravity theory:
\begin{align} \label{ESNCgravity}
    S = \frac{1}{2\kappa^2} \int d^4 x \, \epsilon^{\mu\nu\rho\sigma} \Bigl [
    	  &-\tfrac{1}{2}  \epsilon_{A'\!B'} E_\mu{}^{A'} E_\nu{}^{B'} R_{\rho\sigma}(M)
	 -  \epsilon_{AB} \epsilon_{A'\!B'} \tau_\mu{}^A E_\nu{}^{A'} R_{\rho\sigma}{}^{BB'} (G) \notag \\
	& + \epsilon_{AB} \tau_\mu{}^A m_\nu{}^B R_{\rho\sigma} (J)
	 - \tfrac{1}{2}  \epsilon_{AB} \tau_\mu{}^A \tau_\nu{}^B R_{\rho\sigma} (S)
    \Bigr ]\,.
\end{align}
We note that the non-central extension gauge field $n_\mu$ does not occur in the action. Nevertheless, the action \eqref{ESNCgravity} is invariant under the corresponding gauge transformation, with parameter $\sigma$, under which the gauge field $m_\mu{}^A$ transforms non-trivially, see eq. \eqref{eq:trnsfs}.

Next, we substitute  the expansions eqs.~\eqref{expansion} and \eqref{expparameters} into the relativistic transformation rules eq.~\eqref{eq:gtrnsf} and
eq.~\eqref{eq:gtrnsf2}. This leads to the following equations for the nonrelativistic transformation rules for finite $\omega$:
\begin{subequations}
\begin{align}
    \omega \delta \tau_\mu{}^A + \frac{1}{\omega} \delta m_\mu{}^A & = \omega \lambda \, \epsilon^A{}_B \tau_\mu{}^B + \frac{1}{\omega} \lr \lambda \, \epsilon^A{}_B m_\mu{}^B + \lambda^{AA'} E_{\mu A'} + \epsilon^A{}_B \tau_\mu{}^B \sigma \rr \,, \label{eq:taumtrnsf} \\[.2truecm]
    \delta E_\mu{}^{A'} & = - \lambda_A{}^{A'} \tau_\mu{}^A + \lambda' \epsilon^{A'}{}_{B'} E_\mu{}^{B'}  \,, \label{eq:Emutrnsf} \\[.2truecm]
        \omega^2 \epsilon_{AB} \tau_{[\mu}{}^A \delta \tau_{\nu]}{}^B &=\frac{1}{2}R_{\mu\nu}{}^A(H)\, \sigma^B - \epsilon_{AB} \tau_{[\mu}{}^A D_{\nu]} \sigma^B
 \,,\label{eq:tautrnsf}\\[.2truecm]
       \delta \Omega_\mu &= \p_\mu \lambda' \,,\label{omegamu}\\[.2truecm]
      \delta \Sigma_\mu + \frac{1}{\omega^2} \delta n_\mu & = \p_\mu \lambda + \frac{1}{\omega^2} \lr \p_\mu \sigma + \epsilon_{AB} \lambda^{AA'} \Omega^B{}_{A'} \rr \,, \\[.2truecm]
  \frac{1}{\omega} \delta \Omega_\mu{}^{AA'} & \! = \frac{1}{\omega} \lr \p_\mu \lambda^{AA'} \! - \epsilon^A{}_B \lambda^{BA'} \Sigma_\mu \! - \epsilon^{A'}{}_{B'} \lambda^{AB'} \Omega_\mu \! + \lambda \, \epsilon^A{}_B \Omega_\mu{}^{BA'} \! + \lambda' \, \epsilon^{A'}{}_{B'} \Omega_\mu{}^{AB'} \rr \,, \\[.2truecm]
    \delta \Omega_\mu - \frac{1}{\omega^2} \delta s_\mu & = \p_\mu \lambda' - \frac{1}{\omega^2} \lr \p_\mu\rho + \epsilon_{A'\!B'} \lambda^{A'C} \Omega_\mu{}^{B'}{}_C \rr \,.\label{last}
             \end{align}
\end{subequations}

In principle, it is straightforward to use these equations to solve for the different transformation rules and take the limit $\omega\rightarrow \infty$.
However, there is a subtlety with deriving the nonrelativistic transformation rule of the gauge field $m_\mu{}^A$ under the $\sigma^A$-transformations. One would like to combine eqs.~\eqref{eq:taumtrnsf} and  \eqref{eq:tautrnsf} to achieve this. However, the presence of the curvature term $R_{\mu\nu}{}^A(H)$ forms an obstruction to derive the transformation  (for finite $\omega$) of $\tau_\mu{}^A$ under $\sigma^A$-transformations and this  is needed in eq.~\eqref{eq:taumtrnsf} to derive the transformation of $m_\mu{}^A$ under $\sigma^A$-transformations.  What helps to resolve this issue is that all $R_{\mu\nu}{}^A(H)$ components correspond to  equations of motion of the nonrelativistic action eq.~\eqref{ESNCgravity}.  Assuming that
$R_{\mu\nu}{}^A(H)=0$, we would derive the following transformation rule of $m_\mu{}^A$:
\begin{equation}
\delta m_\mu{}^A  =\partial_\mu\sigma^A -\epsilon^A{}_{B}\sigma^{B}\Sigma_\mu{}\,.
\end{equation}
Since in the off-shell nonrelativistic action we cannot use that $R_{\mu\nu}{}^A(H)=0$\,, this transformation rule will not leave the  action invariant by itself but violate it by terms that are proportional to $R_{\mu\nu}{}^A(H)$. However, because all components of $R_{\mu\nu}{}^A(H)$ are equations of motion, it is guaranteed that we can cancel all these terms by assigning appropriate transformation rules to the fields that give rise to these equations of  motion. A careful analysis  of the nonrelativistic action eq.~\eqref{ESNCgravity} shows that  the fields below have the following transformations under the $\sigma^A$-transformations:
\begin{subequations}\label{rewrite}
\begin{align}
	\delta s_A&=R_{AB}(J) \, \sigma^B\,,
		&
	\delta s_{A'}&= \tfrac{1}{2} R_{A'B}(J) \, \sigma^B\,,\\[.2truecm]
	\delta\Omega_{A'}{}^{AA'}&=- \tfrac{1}{2}\epsilon^{A'\!B'} R_{A'\!B'} (J) \, \sigma^A\,,		&
	\delta \Omega_{(AB)}{}^{A'} -\textrm{trace} & =\epsilon^{A'B'} R_{B'(A}(J) \, \sigma_{B)}-\textrm{trace}\,,\\[.2truecm]
	\delta \Omega_{[AB]}{}^{A'}&=\epsilon^{A'B'} R_{B'[A}(J) \, \sigma_{B]}\,,
		&
	\delta\Sigma_A & = - \tfrac{1}{2} \epsilon_{AB}\,\delta\Omega_{A'}{}^{BA'}\,,
\end{align}
\end{subequations}
where the trace is taken over the longitudinal $(AB)$ indices.  We have used here the nonrelativistic  inverse fields
$\tau^\mu{}_A$ and $E^\mu{}_A$\,, which are defined by the following relations:
\begin{eqnarray}\label{inverses}
E_{\mu}{}^{A^\prime} E^{\mu}{}_{B^\prime} & = &\delta_{B^\prime}^{A^\prime}, \ \ \ \ \ E_{\mu}{}^{A^\prime} E^{\nu}{}_{A^\prime}  = \delta^{\nu}_{\mu} - \tau_{\mu}{}^A\tau^{\nu}{}_A\,,\ \ \ \ \ \tau^{\mu}{}_A\tau_{\mu}{}^B = \delta_A^B\,,\nonumber\\[.2truecm]
\tau^{\mu}{}_A E_{\mu}{}^{A^\prime} & = &0, \ \ \ \ \ \ \ \tau_{\mu}{}^A E^{\mu}{}_{A^\prime} =  0\,.
\end{eqnarray}
Furthermore, we have used the following simplified notation: when a spacetime index is contracted with the spacetime index of an (inverse) Vielbein field (with no derivatives acting on it), we replace this spacetime index with the flat index of this (inverse) Vielbein field. For example,
\begin{equation}
R_{AB} (J) = \tau^{\mu}{}_{A} \tau^{\nu}{}_{B} R_{\mu\nu} (J)\,.
\end{equation}
We  can further rewrite the equations \eqref{rewrite}
as
\begin{subequations}
\begin{align}
	\delta \Sigma_\mu & = - \tfrac{1}{2} \, \tau_{\mu}{}^{A} \sigma^B \epsilon_{AB} R (J)\,, \\[2pt]
	\delta s_\mu & = \tfrac{1}{2} \ls \tau_\mu{}^A \sigma^B R_{AB} (J) + \sigma^B R_{\mu B} (J) \rs \,, \\[2pt]
	\delta \Omega_\mu{}^{AA'} & = \tau_\mu{}^B \epsilon^{A'\!B'} \ls \sigma^A R_{B'\!B} (J) - \tfrac{1}{2} \sigma^C \delta^{A}_{B} \, R_{B'\!C} (J) \rs + \tfrac{1}{2} E_\mu{}^{A'} \sigma^A R (J) \,,
\end{align}
\end{subequations}
where we have defined $R_{A'\!B'} (J) = \epsilon_{A'\!B'} R(J)$.

Having resolved this subtlety,
one obtains, after taking the limit $\omega \rightarrow \infty$,  precisely the nonrelativistic transformation rules eq.~\eqref{eq:trnsfs} derived in the previous section\,\footnote{Except for the $\sigma^A$-transformations, see above.} with all fields independent.\,\footnote{We cannot derive in this way the transformation rule of the non-central extension gauge field $n_\mu$. This transformation rule will not be needed since the action eq.~\eqref{ESNCgravity} is independent of $n_\mu$.} As a consistency check one may verify that the action eq.~\eqref{ESNCgravity} is invariant under these nonrelativistic transformations.

The action eq.~\eqref{ESNCgravity} provides a first-order formulation of the 4D ESNC gravity theory. In order to go to a second-order formulation we consider those equations of motion that give rise to conventional constraints on the curvatures. For this, we only need to vary the spin-connection fields $\Sigma_\mu, \Omega_\mu{}^{AA'}$ and $\Omega_\mu$  in the first three terms. We do not vary  the central charge gauge field $s_\mu$ in the last term since that variation would lead to the 4 geometric constraints
\begin{equation}\label{foliation2}
	\epsilon^{}_{AB} \tau_{[\mu}{}^A \partial^{\phantom{p}}_{\nu} \tau_{\rho]}{}^B = 0\ \ \ (2+2)\hskip .5truecm \textrm{or}\hskip .5truecm
\tau_{A^\prime A}{}^A  =0\ \ \ (2)\hskip .5truecm \textrm{and}\hskip .5truecm  \tau_{A^\prime B^\prime A}=0\ \ \ (2)\,,
\end{equation}
where we have defined
\begin{equation}
\tau^{}_{\mu\nu}{}^A \equiv \p^{}_{[\mu} \tau^{}_{\nu]}{}^A\,.
\end{equation}
The numbers in brackets in eq.~\eqref{foliation2} indicate the number of constraints.

As it turns out, not all components of $\Omega_\mu{}^{AA'}$ lead to conventional  constraints. These are precisely the components that remain independent and cannot be solved for in terms of the Vierbein fields. To be concrete, the variation of the 4 components defined by
\begin{equation} \label{eq:WABBA'def}
W_{AB}{}^{A'} \equiv  \Omega_{(A B)}{}^{A'} - \textrm{trace} \ \ \ (4)
\end{equation}
 leads to the 4 geometric constraints
\begin{equation}\label{geometric1}
\tau_{A'(AB)} - \textrm{trace} =0\ \ \ (4)\,.
\end{equation}
We will therefore not vary these 4 components. Finally, combining eqs.~\eqref{foliation2} and \eqref{geometric1}, we obtain in total 8 geometric constraints that can be written as follows:
\be \label{eq:geometricconstraints}
	\tau_{A'(AB)} = 0 \ \ \  (6)\,,
		\qquad
	\tau_{A'B'A} = 0 \ \ \ (2)\,.
\ee

Varying the remaining 20 components of the spin-connection fields  $\Sigma_\mu (4)\,, \Omega_\mu (4)$ and $\Omega_\mu{}^{AA'} (16-4=12)$,  we obtain the following constraints:
\begin{subequations} \label{eq:othercon}
\begin{align}
    \epsilon_{A'\!B'} E_{[\mu}{}^{A'} R_{\nu\rho]}{}^{B'}(P)  & = 0 \quad (4) \,, \\[2pt]
    \epsilon_{AB} \ls m_{[\mu}{}^A R_{\nu\rho]}{}^B (H) + \tau_{\mu}{}^A R_{\nu\rho}{}^B (Z) \rs & = 0 \quad (4) \,, \\[2pt]
    E_{[\mu}{}^{A'} R_{\nu\rho]}{}^A (H) - \tau_{[\mu}{}^A R_{\nu\rho]}{}^{A'} (P)
    - `\textrm{projection'}& = 0 \quad (12)\,.\label{33c}
\end{align}
\end{subequations}
By `projection' in eq.~\eqref{33c} we mean that the field equations that correspond to the variation of $W_{AB}{}^{A'}$ are not included.\,\footnote{
These components are most easily identified by first multiplying  equation \eqref{33c} with $\epsilon^{\sigma\mu\nu\rho}$ so that the free indices are $\sigma, A$ and $A'$. One next converts the free $\sigma$-index into a free flat index $B$ such that the free flat indices are $A, B$ and $B'$. Finally, one subtracts the trace in the longitudinal indices $A$ and $B$.}
The above 20 constraints are all conventional and can be used to solve for the following 20 spin-connection components:
\begin{subequations}\label{extraterms}
\begin{align}
	\Sigma^{}_\mu & = \epsilon^{AB} \lr \tau^{}_{\mu AB} - \tfrac{1}{2} \tau^{}_\mu{}^C \tau^{}_{ABC}\rr + \Delta_\mu(\Sigma) & & (4)\,, \\[4pt]
	\Omega^{}_\mu & = \epsilon^{A'\!B'} \lr - E^{}_{\mu A'\!B'} + \tfrac{1}{2} E^{}_{\mu}{}^{C'} E^{}_{A'B'C'}  + \tfrac{1}{2} \tau^{}_\mu{}^A \Omega^{}_{A'\!AB'} \rr
+\Delta_\mu(\Omega)& & (4)\,, \\[4pt]
\Omega^{}_\mu{}^{AA'} & = - E^{}_\mu{}^{AA'} + E^{}_{\mu B'} E^{AA'\!B'} + m_\mu{}^{A'\!A} + \tau^{}_{\mu B} m^{AA'\!B} + \tau_{\mu B} \tilde{W}^{ABA'} + \Delta_\mu{}^{AA'}(\Omega)& & (12)\,,	\label{eq:OmegaDeltaAA'}
\end{align}
\end{subequations}
where the components $\tilde{W}^{ABA'}$ are related to the components $W_{AB}{}^{A'}$ defined in eq.~\eqref{eq:WABBA'def} as follows:
\be
	\tilde{W}_{AB}{}^{A'} \equiv W_{AB}{}^{A'} + 2 \bigl(m^{A'\!}{}_{(AB)} - \text{trace}\bigr) - \bigl(\Delta_{(AB)}{}^{A'}(\Omega) - \text{trace}\bigr)\,.
\ee
In the above equations we have used the definitions
\be \label{definitions}
	E^{}_{\mu\nu}{}^A \equiv \p^{}_{[\mu} E^{}_{\nu]}{}^{A'}\,,
		\qquad \qquad
	m^{}_{\mu\nu}{}^A \equiv \p^{}_{[\mu} m^{}_{\nu]}{}^A + \epsilon^A{}^{}_B m^{}_{[\mu}{}^B \Sigma^{}_{\nu]}\,.
\ee
The $\Delta$-terms are torsion contributions to the spin-connections given by
\begin{subequations}
\begin{align}
 \Delta_\mu(\Sigma) &=0   \,, \\[4pt]
 \Delta_\mu(\Omega) &= \tfrac{1}{2}\epsilon^{A'\!B'} R_{A'\!A}{}^A (H) E_{\mu B'} \,, \\[4pt]
 \Delta_\mu{}^{AA'}(\Omega) &=\tfrac{1}{2}\tau_{\mu B} \lr 2 R^{AA'[B}(H)m_C{}^{C]} + \eta^{AB} R_C{}^{A'[D}(H) m{}_D{}^{C]} + R_{CA'}{}^C(H) m^{[AB]} \rr \notag \\[2pt]
 	& \qquad  +   R_{\mu A'}{}^{[A}(H) m_B{}^{B]} - R_{\mu B}{}^{[A}(H) m^{|A'|B]} - R_B{}^{A'[A}(H) m_\mu{}^{B]}  \,.
 \end{align}
 \end{subequations}
 These torsion terms will vanish once the geometric constraints are imposed.
Substituting the expressions eq.~\eqref{extraterms} for the spin-connections back into the action eq.~\eqref{ESNCgravity} we obtain the following second-order formulation of the 4D ESNC gravity theory:
\begin{align} \label{second-order}
    S = \frac{1}{2\kappa^2} \int d^4 x \, \epsilon^{\mu\nu\rho\sigma} \Bigl [
    	  &-\tfrac{1}{2}  \epsilon_{A'\!B'} E_\mu{}^{A'} E_\nu{}^{B'} R_{\rho\sigma}(M)
	 -  \epsilon_{AB} \epsilon_{A'\!B'} \tau_\mu{}^A E_\nu{}^{A'} R'_{\rho\sigma}{}^{BB'} (G)
	 + \epsilon_{AB} \tau_\mu{}^A m_\nu{}^B R_{\rho\sigma} (J)\notag\\[.2truecm]
	& - 2 \epsilon_{AB} \epsilon_{A'\!B'} \tau_\mu{}^A E_\nu{}^{A'} \partial_\rho(\tau_{\sigma C} W^{BCB'}) - \tfrac{1}{2}  \epsilon_{AB} \tau_\mu{}^A \tau_\nu{}^B R_{\rho\sigma} (S)
    \Bigr ]\,.
\end{align}
It is now understood that all three spin-connection fields are dependent and given by the expressions eq.~\eqref{extraterms}. The prime in the boost curvature $R(G)$ indicates that we have left out the independent spin-connection components $W_{AB}{}^{A'}$. They appear as a separate term in the second line of eq.~\eqref{second-order}.

Finally, given the above second-order action eq.~\eqref{second-order}, the equations of motion can be obtained by varying the independent fields $\{\tau_\mu{}^A, E_\mu{}^{A'}, m_\mu{}^A\}$ together with the independent Lagrange multipliers $\{W_{AB}{}^{A'}, s_\mu\}$. The Lagrange multipliers occur in the second line of eq.~\eqref{second-order} and give rise to the 8 geometric constraints given in eq.~\eqref{eq:geometricconstraints}. As a consequence, all $\Delta$-terms in eq.~\eqref{extraterms} vanish. Using the resulting  simplified expressions for the spin-connections, the equations of motion corresponding to the independent fields  $\{\tau_\mu{}^A, E_\mu{}^{A'}, m_\mu{}^A\}$ are respectively given by
\begin{subequations}
\begin{align}
    \epsilon_{A'\!B'} E_{[\mu}{}^{A'} R_{\rho\sigma]}{}^{AB'} (G) - m_{[\mu}{}^A R_{\nu\rho]} (J) + \tau_{[\mu}{}^A R_{\nu\rho]}(S) & = 0\,, \\[4pt]
	E_{[\mu}{}^{A'} R_{\nu\rho]} (M) - \epsilon^{}_{AB} \tau_{[\mu}{}^A R_{\nu\rho]}{}^{BA'} (G) & = 0\,, \\[4pt]
	\tau_{[\mu}{}^A R_{\nu\rho]} (J) & = 0\,.
\end{align}
\end{subequations}
The above equations are a generalization of the usual Newton-Cartan equations of motion. They do not describe any momentum mode that could replace the graviton of general relativity.  Instead, they describe the (instantaneous) gravitational interaction between massive objects. For instance, in the case of string NC gravity, similar equations describe the instantaneous gravitational attraction between the wound strings that occur in the spectrum of a non-relativistic string \cite{Gomis:2000bd}.

This concludes our discussion of the 4D ESNC gravity theory.

\section{Relation to 3D Extended \NC Gravity} \label{sec:3D}

In this section we discuss the relation between the present work and earlier work in three dimensions \cite{Bergshoeff:2016lwr, Hartong:2016yrf}.
By taking a dimensional reduction along the longitudinal spatial direction ($A=1$) followed by a truncation setting some of the generators equal to zero, see eq.~\eqref{eq:below}, the 4D ESNC algebra eq.~\eqref{eq:ESNCa} reduces to the 3D algebra that underlies the 3D extended NC gravity \cite{Bergshoeff:2016lwr, Hartong:2016yrf}.

The dimensional reduction and truncation can be done by using the following prescriptions:
\begin{align}
	\eta_{AB} \rightarrow
		\begin{pmatrix}
			-1\,\, & 0 \,\, \\
			0\,\, & 0 \,\,
		\end{pmatrix}\,,
\end{align}
and
\begin{subequations} \label{eq:below}
\begin{align}\label{below1}
	H_0 & \rightarrow \bar H\,,
		&
	H_1 & \rightarrow 0\,,
		&
	P_{A'} & \rightarrow \bar P_{A'}\,; \\[5pt]
	G_{0A'} & \rightarrow \bar G_{A'}\,,
		&
	G_{1A'} & \rightarrow  0\,,
		&
	M & \rightarrow 0\,,\label{below2}
		&
	J & \rightarrow \bar J\,; \\[5pt]
	Z & \rightarrow 0\,,
		&
	Z_0 & \rightarrow \bar Z\,,
		&
	Z_1 & \rightarrow 0\,,
		&
	S & \rightarrow - \bar S\,.\label{below3}
\end{align}
\end{subequations}
where the barred generators correspond to the symmetries of the reduced 3D algebra as follows:
\begin{align*}
    \text{time  translations} \qquad & \bar H \\
    \text{transverse translations} \qquad & {\bar P}_{A'} \\
    \text{Galilei boosts} \qquad & {\bar G}_{A'} \\
    \text{spatial rotations} \qquad & \bar J \\
    \text{central extensions} \qquad & \bar Z,\, \bar S
\end{align*}
We use the following  terminology for the algebras that are formed by different sets of the above generators:
\begin{enumerate}

\item

	\emph{Galilei algebra} consists of the generators $\bar{H}\,, \bar{P}_{A'}\,, \bar{G}_{A'}$ and $\bar{J}$\,.
	
\item

	\emph{Newton-Cartan algebra} is a central extension of the Galilei algebra that includes the generator $\bar{Z}$\,. This algebra underlies NC gravity. In the literature, this NC algebra is usually referred to as the Bargmann algebra.

\item

	\emph{Extended Newton-Cartan algebra} is a central extension of the 3D NC algebra that includes the generator $\bar{S}$\,. This algebra underlies extended NC gravity. Note that in \cite{Bergshoeff:2016lwr} a different terminology is used. There, the extended NC algebra is referred to as the extended Bargmann algebra and the extended NC gravity is referred to as the extended Bargmann gravity.

\end{enumerate}
We emphasize that the Galilei and NC algebras exist in any dimension, but that the extended NC algebra only exists in three dimensions.

As a result, the ESNC algebra eq.~\eqref{eq:ESNCa} reduces to
\begin{subequations}\label{eq:EBa}
\begin{align}
	[\bar G_{A'}\,, \bar H] & = \bar P_{A'}\,, &
	[\bar P_{A'}\,, \bar J]  & = \epsilon_{A'}{}^{B'} \bar P_{B'}\,, &
	[\bar G_{A'}\,, \bar J]  & = \epsilon_{A'}{}^{B'} \bar G_{B'}\,; \\[2pt]
	[\bar G_{A'}\,, \bar P_{B'}] & = \delta_{A'\!B'} \bar Z\,, &
	[\bar G_{A'}\,, \bar G_{B'}]  & = \epsilon_{A'B'} \bar S\,. \label{eq:EBaGG}
\end{align}	
\end{subequations}
This matches the extended NC algebra introduced in \cite{Bergshoeff:2016lwr, Hartong:2016yrf}.

We can perform a similar dimensional reduction and truncation of the action eq.~\eqref{ESNCgravity} along the longitudinal spatial $y$-direction which we take to be an  isometry direction. Using  the adapted coordinate system  $x^\mu = (x^i\,, y)$\,, $i = 0\,, 2\,, 3$ we have the following reduction and truncation of the various fields:
\begin{subequations}
\begin{align}
	E_y{}^{A'} & \rightarrow 0\,,
		&
	m_i{}^0 & \rightarrow \bar m_i\,,
		&
	m_i{}^1 \,,\,
	m_y{}^A & \rightarrow 0\,; \\[2pt]
	\tau_y{}^1 & \rightarrow 1\,,
		&
	\tau_i{}^0 & \rightarrow \bar\tau_i\,,
		&
	\tau_i{}^1\,,\,
	\tau_y{}^0 & \rightarrow 0\,.
\end{align}
\end{subequations}
The gauge field $s_\mu$ reduces to
\be
	s_y \rightarrow 0\,,
		\qquad
	s_i \rightarrow - \bar s_i\,.
\ee
Moreover, the spin connections reduce to
\begin{align}
	\Omega_i \rightarrow \bar \Omega_i\,,
		\qquad
	\Omega_i{}^{0A'} \rightarrow \bar\Omega_i{}^{A'}\,,
		\qquad
	\Sigma_\mu\,, \,\,\Omega_y\,, \,\,\Omega_y{}^{0A'}, \,\,\Omega_\mu{}^{1A'} \rightarrow 0\,.
\end{align}
From this it follows that the various curvature two-forms reduce to
\begin{subequations}
\begin{align}
	R_{ij} (M) & \rightarrow 0\,, \\[5pt]
	R_{ij} (J) & \rightarrow \bar{R}_{ij} (\bar J) = 2 \lr \p_{[i} \bar E_{j]}{}^{A'} + \epsilon^{A'}{}_{B'} \bar{E}_{[i}{}^{B'} \bar\Omega_{j]} + \bar\tau_{[i} \bar\Omega_{j]}{}^{A'} \rr\,, \\[5pt]
	R_{ij}{}^{0A'} (G) & \rightarrow \bar{R}_{ij}{}^{A'} (\bar G) = 2 \lr \p_{[i} \bar\Omega_{j]}{}^{A'} + \epsilon^{A'}{}_{B'} \bar\Omega_{[i}{}^{B'} \bar\Omega_{j]} \rr\,, \\[5pt]
	R_{ij}{}^{1A'} (G) & \rightarrow 0\,, \\[5pt]
	R_{ij} (S) & \rightarrow - \bar{R}_{ij} (\bar S) = - 2 \p_{[i} \bar s_{j]} + \epsilon_{A'\!B'} \bar\Omega_{[i}{}^{A'} \bar\Omega_{j]}{}^{B'}\,.
\end{align}
\end{subequations}
Note that all curvature two-tensors with $y$ indices vanish after dimensional reduction. It then follows that the 4D ESNC gravitational action eq.~\eqref{ESNCgravity} dimensionally reduces to
\begin{align}
	S = \frac{1}{2\kappa^2} \int dy \int d^3 x \, \epsilon^{yijk} \Bigl [ \epsilon_{A'\!B'} E_i{}^{A'} \bar{R}_{jk}{}^{B'} (\bar G) - \bar{m}_i \bar{R}_{jk}(\bar J) - \bar\tau_i \bar{R}_{jk} (\bar S) \Bigr ]\,.
\end{align}
Hence, we find that the 3D action is given by
\begin{align} \label{eq:EBaction}
	S = \frac{k}{4\pi} \int d^3 x \, \epsilon^{ijk} \Bigl [ \epsilon_{A'\!B'} E_i{}^{A'} \bar{R}_{jk}{}^{B'} (\bar G) - \bar{m}_i \bar{R}_{jk}(\bar J) - \bar\tau_i \bar{R}_{jk} (\bar S) \Bigr ]\,,
\end{align}
where $k$ is the Chern-Simons level.

To match with the conventions in \cite{Bergshoeff:2016lwr}, one needs to redefine the Galilean boost generator $G_{A'}$ as
\be
	\bar G_{A'} \rightarrow - \epsilon_{A'}{}^{B'} \bar G_{B'}\,.
\ee
Then, the commutation relations eq.~\eqref{eq:EBa} become
\begin{subequations}
\begin{align}
	[\bar H\,, \bar G_{A'}] & = - \epsilon_{A'}{}^{B'} \bar P_{A'}\,, \hskip 1truecm
	[\bar J\,, \bar P_{A'}]  = - \epsilon_{A'}{}^{B'} \bar P_{B'}\,, \hskip 1truecm
	[\bar J\,, \bar G_{A'}]  = - \epsilon_{A'}{}^{B'} \bar G_{B'}\,, \\[10pt]
	[\bar G_{A'}\,, \bar P_{B'}] & = \epsilon_{A'\!B'} \bar Z\,, \hskip 2truecm
	[\bar G_{A'}\,, \bar G_{B'}]  = \epsilon_{A'B'} \bar S\,.
\end{align}	
\end{subequations}
Furthermore, we need to redefine
\be
	\bar\Omega_i{}^{A'} \rightarrow - \epsilon^{A'}{}_{B'} \bar\Omega_i{}^{B'}\,.
\ee
Consequently,
\be
	\bar{R}_{ij}{}^{A'} (\bar G) \rightarrow - \epsilon^{A'}{}_{B'} \bar{R}_{ij}{}^{A'} (\bar G)\,,
\ee
where the definition of $\bar{R}_{ij}{}^{A'}(\bar G)$ does not change. From all this it follows that the action  eq.~\eqref{eq:EBaction} now reads
\begin{align}
	S = \frac{k}{4\pi} \int d^3 x \, \epsilon^{ijk} \Bigl [ E_i{}^{A'} \bar{R}_{jkB'} (\bar G) - \bar{m}_i \bar{R}_{jk}(\bar J) - \bar\tau_i \bar{R}_{jk} (\bar S) \Bigr ]\,,
\end{align}
which precisely equals  eq.~(5) in \cite{Bergshoeff:2016lwr}. Here, the Levi-Civita symbol $\epsilon^{ijk}$ is defined such that $\epsilon^{012} = 1$\,.

Before closing this section, we mention briefly some further extensions of the ESNC algebra eq.~\eqref{eq:ESNCa} that one might consider.
First, we introduce an extra generator $S_{AB}$ with $S_{AB} = S_{BA}$ and $\eta^{AB} S_{AB} = 2 S$ with $S$ being the central charge generator we considered in this work.  Then, the commutation relation eq.~\eqref{eq:ESNCaGG} is modified to be
\be \label{eq:extendedESNCGG}
	[G_{AA'}\,,\, G_{BB'}] = \delta_{A'\!B'} \epsilon_{AB} Z + \epsilon_{A'\!B'} S_{AB}\,.
\ee
In order to close the algebra, one has to include a new commutation relation,
\be \label{eq:extendedESNCSM}
	[S_{AB}\,,\, M] = \epsilon_A{}^C S_{BC} + \epsilon_B{}^C S_{AC}\,.
\ee
Second, in addition to $S_{AB}$\,, one may also include another generator $S_A$ by modifying eq.~\eqref{eq:ESNCaGP} to be
\be \label{eq:extendedESNCGP}
	[G_{AA'}\,,\, P_{B'}] = \delta_{A'\!B'} Z_A + \epsilon_{A'\!B'} S_A\,.
\ee
In order to close the algebra, one has to include two new commutation relations,
\begin{subequations} \label{eq:extendedESNCSMHS}
\begin{align}
	[S_A\,,\, M] & = \epsilon_A{}^B S_B\,, \\
	[H_A\,,\, S_{BC}] & = \eta_{AB} S_C + \eta_{AC} S_B\,.
\end{align}
\end{subequations}

In order to dimensionally reduce to three dimensions this new algebra that combines eqs.~\eqref{eq:ESNCa} and \eqref{eq:extendedESNCGG} $\sim$ \eqref{eq:extendedESNCGP}, we take
\be
	S_0 \rightarrow - \bar Y\,,
		\qquad
	S_1 \rightarrow 0\,,
	 	\qquad
	S_{AB} \rightarrow - \eta_{AB} \bar{S}\,.
\ee
Then, the commutation relations eq.~\eqref{eq:EBaGG} become
\be
	[\bar G_{A'}\,,\, \bar P_{B'}] = \epsilon_{A'\!B'} \bar{Z} - \epsilon_{A'\!B'} \bar{Y}\,,\hskip 2truecm
[\bar G_{A'}\,,\, \bar G_{B'}] = \epsilon_{A'\!B'} \bar S\,.
\ee
In addition, we have one other non-vanishing commutator,
\be
	[\bar H\,,\, \bar{S}] = 2 \bar{Y}\,.
\ee
Together, these commutation relations give the algebra described in \cite{Hartong:2016yrf}.

\section{Conclusions} \label{sec:concl}

In this work we showed how an action can be constructed for the 4D string Newton-Cartan (NC) gravity theory underlying nonrelativistic string theory provided an extra gauge field is introduced that signals a central extension in the string NC algebra. We called this extended gravitational theory Extended String Newton-Cartan (ESNC) gravity. The extra vector field is on top of the zero-flux two-form gauge field that needs to be added to obtain a finite limit of the relativistic string worldsheet action \cite{Gomis:2000bd}. The vector and two-form gauge fields occur together in an extra term that needs to be added to the Einstein-Hilbert term in order to obtain a finite limit of the relativistic target space action, see eq.~\eqref{eq:relaction}. It would be interesting to see whether, requiring conformal invariance of the worldsheet action of 4D nonrelativistic string theory, leads to the equations of motion of string NC gravity (without an action) or to its ESNC gravity version (with an action) that is discussed in this work.

Our results are a natural extension of earlier work in three dimensions \cite{Bergshoeff:2016lwr,Hartong:2016yrf} where a Chern-Simons (CS) action was constructed for the extended (particle) NC algebra. Indeed, we have shown how the extended NC algebra can be obtained from the ESNC algebra by reduction and  truncation and how the 3D extended NC gravity action can be obtained from the 4D ESNC gravity action by a reduction over the spatial longitudinal direction followed by a truncation. It is clear that our procedure for constructing a nonrelativistic action can be generalized to general relativity in $p+3$ dimensions where  the divergences originating from the Einstein-Hilbert term are cancelled by adding the following term containing a ($p+1$)-form gauge field $\hat{\CB}_{\mu_1\cdots \mu_{p+1}}$ and a one-form gauge field $\hat{\CA}_{\mu}$:
\begin{equation}\label{secondtermp+3}
\epsilon^{\mu_1\cdots \mu_{p+3}} \hat{\CB}_{\mu_1\cdots \mu_{p+1}}\partial_{\mu_{p+2}}\hat{\CA}_{\mu_{p+3}}\,.
\end{equation}
The nonrelativistic limit leads to an action for  {\sl extended $p$-brane NC gravity}. The resulting nonrelativistic actions  in different dimensions are related to each other by  a dimensional reduction over one of the spatial  isometry directions longitudinal to the $p$-brane followed by a truncation. It has been shown that an alternative way to obtain  the 3D extended NC gravity action is to start from the Einstein-Hilbert action and to apply the technique of so-called Lie algebra expansions \cite{private}. It would be interesting to see whether, similarly,  the higher-dimensional $p$-brane NC gravity actions can be obtained by appropriate generalizations of such Lie algebra expansions.

The 4D action constructed in this work is an extension of a 3D CS action in the sense that all terms in the first-order action eq.~\eqref{ESNCgravity} are the wedge product of \emph{two} one-form gauge fields with one two-form curvature as opposed to a 3D CS action where all terms are wedge  products of a \emph{single} one-form gauge field with a two-form curvature. The necessary condition for writing down a 3D CS action is that the underlying algebra allows a non-degenerate symmetric invariant bilinear form.
The ESNC algebra does not possess such a bilinear form (see appendix \ref{app:form}).
Evidently, the non-existence of a non-degenerate symmetric invariant bilinear form for the ESNC algebra is not an obstruction to writing down an action of the form eq.~\eqref{ESNCgravity}. Instead, such an action requires the existence of a particular trilinear form. Requiring the action to be gauge-invariant already leads to constraints on this trilinear form. Another difference with the 3D case is that the 4D action is not invariant under the P-translations of the algebra whereas in 3D the P-translations are on-shell equivalent to the general coordinate transformations. It would be interesting to determine the minimum set of requirements that must be met at the algebraic level in order to be able to write down a nonrelativistic gravity action
of the form eq.~\eqref{ESNCgravity}.

Having constructed an action for nonrelativistic gravity  is one step further in investigating the implications of nonrelativistic holography with nonrelativistic gravity in the bulk \cite{Bagchi:2009my}. The price we had to pay in order to write down an action  was the introduction of an extra vector field corresponding to a central extension in the algebra. It would be interesting to see how this extra vector field would fit into the target space geometry of 4D nonrelativistic string theory or its variation. In the case of the 3D (particle) NC algebra, the central extension has been related to the occurrence of anyons in three dimensions
\cite{Duval:2000xr,Jackiw:2000tz}. What makes anyons possible in three dimensions is the fact that the transverse rotation group of a 3D particle is Abelian. The same property has been used to show that the spectrum analysis of 3D relativistic string theory does not follow  the general pattern in arbitrary dimensions but, instead, gives rise to anyonic particles in the spectrum \cite{Mezincescu:2010yp}. Here, in the nonrelativistic case, we are facing a similar situation with nonrelativistic strings  in four dimensions where the rotation group transverse to the string is Abelian.
It would be interesting to see whether the general spectrum analysis of nonrelativistic string theory \cite{Gomis:2000bd} becomes special in four dimensions and could lead to extra (winding) anyonic strings in the spectrum that are absent in the general analysis. Such extra states could have  important implications for the  3D boundary field theory such as the fractional quantum Hall effect where anyons do play a role. For a recent discussion of anyonic strings in the bulk and anyons/vortices at the boundary within a holographic context, see
\cite{Gussmann:2018leh}.

\section*{Acknowledgements}
We thank Jaume Gomis, Troels Harmark, Jelle Hartong, Lorenzo Menculini, Djordje Minic, Niels Obers and Jan Rosseel for useful discussions. In particular, we wish to thank Jan Rosseel for pointing out the relationship between this work and constrained BF theory.
K.T.G. and Z.Y. are grateful for the hospitality of the University of Groningen, where this work was initiated; they also thank Michael Thisted for his generosity in hosting them in Skanderborg, Denmark, where part of this work was done.
Z.Y. is also grateful for the hospitality of Julius-Maximilians-Universit\"{a}t W\"{u}rzburg and the Niels Bohr Institute.
The work of K.T.G. was supported in part by the Free Danish Research Council (FNU) grant ``Quantum Geometry" and the Independent Research Fund Denmark project ``Towards a deeper understanding of black holes with nonrelativistic holography" (grant number DFF-6108-00340).
The work of C.\c{S}. is part of the research programme of the Foundation for Fundamental Research on Matter (FOM), which is financially supported by the Netherlands Organisation for Science Research (NWO).
This research was supported in part by Perimeter Institute
for Theoretical Physics. Research at Perimeter Institute is supported by the
Government of Canada through the Department of Innovation, Science and Economic
Development and by the Province of Ontario through the Ministry of Research, Innovation
and Science.

\appendix

\section{Irreducible Gauging} \label{app:irreducible}

To obtain an irreducible gauge theory (without an action) we impose the following maximum set of 36 conventional constraints on the curvatures:
\begin{subequations}
\begin{align}
	R_{\mu\nu}{}^A(H)  &=  0 \ \ \ (12)\,,\label{constraints1}\\[.1truecm]
	R_{\mu\nu}{}^{A^\prime}(P)   &= 0\ \ \ (12)\,,\label{constraints2}\\[.1truecm]
	R_{\mu\nu}{}^A(Z)  &=0\ \ \   (12)\,,\label{constraints3}
\end{align}
\end{subequations}
where we have indicated the number of constraints between brackets.
The first 12 constraints are special in the sense that they are a mixture of 4 conventional constraints that are  used to solve for the 4 components of the longitudinal spin-connection $\Sigma_\mu$ and the
 8 geometric  constraints given in eq.~\eqref{eq:geometricconstraints}.

The remaining $24 = 12+12$ constraints in eqs.~\eqref{constraints2} and \eqref{constraints3} are used to solve for the 16 Galilean boost spin connections $\Omega_\mu{}^{AA^\prime}$, the 4 transverse spin-connections $\Omega_\mu$ and the 4 non-central extension fields $n_\mu$ that did not occur in the nonrelativistic action.
The calculation of the explicit expressions of the dependent gauge fields
\be
\{\Sigma_\mu\,, \Omega_\mu{}^{AA^\prime}\,, \Omega_\mu\,, n_\mu\}
\ee
is identical to the one corresponding to the string NC algebra \cite{inpreparation} since the extra generator $S$, being a central extension, does not occur in the curvatures that  are set to zero. For the convenience of the reader we give the resulting expressions \cite{inpreparation},\,\footnote{To compare with \cite{inpreparation}, one should use the following dictionary:
\be
	\Omega_\mu{}^{AB} \rightarrow \epsilon^{AB} \Sigma_\mu\,,
		\qquad%
	\Omega_\mu{}^{A'\!B'} \rightarrow \epsilon^{A'\!B'} \Omega_\mu\,,
		\qquad%
	n_\mu{}^{ab} \rightarrow - \epsilon^{AB} n_\mu\,.
\ee}
\begin{subequations}
\begin{align}
	\Sigma^{}_\mu & = \epsilon^{AB} \lr \tau^{}_{\mu AB} - \tfrac{1}{2} \tau^{}_\mu{}^C \tau^{}_{ABC} \rr\,, \\[4pt]
	\Omega^{}_\mu & = \epsilon^{A'\!B'} \lr - E^{}_{\mu A'\!B'} + \tfrac{1}{2} E^{}_{\mu}{}^{C'} E^{}_{A'B'C'}  + \tfrac{1}{2} \tau^{}_\mu{}^A \Omega^{}_{A'\!AB'} \rr\,, \\[4pt]
	\Omega^{}_\mu{}^{AA'} & = - E^{}_\mu{}^{AA'} + E^{}_{\mu B'} E^{AA'\!B'} + m_\mu{}^{A'\!A} + \tau^{}_{\mu B} m^{AA'\!B}\,, \\[4pt]
	n^{}_\mu{} & = \epsilon^{AB} \lr m^{}_{\mu AB} - \tfrac{1}{2} \tau^{}_\mu{}^C m^{}_{ABC} - \tfrac{1}{2} E^{}_{\mu}{}^{A'} E^{}_{ABA'} \rr\,.
\end{align}
\end{subequations}
where we have used the definitions eq.~\eqref{definitions}.
The transformations of these dependent gauge fields are identical to the ones given in eq.~\eqref{eq:trnsfs} before imposing the curvature constraints,
except for an additional longitudinal Lorentz curvature term in the transformation of the $n_\mu$ gauge-field under a gauge transformation with parameter $\sigma^A$ \cite{inpreparation}. This is due to the fact that the additional constraint $R_{\mu\nu}{}^A(Z) = 0$  that we impose is not invariant under $\sigma^A$-transformations, but instead transform into a longitudinal Lorentz curvature term.
 The explicit form of this term will not be needed since the gauge field $n_\mu$ does not occur in the nonrelativistic action eq.~\eqref{ESNCgravity}.

\section{Extended $p$-Brane Newton-Cartan Algebras} \label{app:pbrane}

We consider generalizations of the ESNC algebra eq.~\eqref{eq:ESNCa} to $p$-branes in $p+3$ dimensions. Take an index $A\, (A = 0, 1, \cdots, p)$ longitudinal to the brane and an index $A'\, (A' = p+1, p+2)$ transverse to the brane. The \emph{extended $p$-brane Newton-Cartan algebra} consists of the generators,
\begin{align*}
    \text{longitudinal translations} \qquad & H_A \\
    \text{transverse translations} \qquad & P_{A'} \\
    \text{longitudinal Lorentz transformations} \qquad & M_{AB} \\
    \text{string Galilei boosts} \qquad & G_{AA'} \\
    \text{spatial rotations} \qquad & J \\
    \text{non-central extensions} \qquad & Z_A\ \textrm{and} \ Z_{AB} \\
    \text{central extension} \qquad & S
\end{align*}
among whom all non-zero commutators are given by
\footnote{See \cite{Brugues:2006yd} for the $p$-brane NC algebra in general dimensions, in the absence of the central charge generator $S$.}
\begin{subequations}
\begin{align}
	[H_A\,, M_{BC}] & = - \eta_{AB} H_C + \eta_{AC} H_B\,, &
	[H_A\,, G_{BA'}] & = \eta_{AB} P_{A'}\,, \\
	[P_{A'}\,, J] & = \epsilon_{A'}{}^{B'} P_{B'}\,, &
	[G_{AA'}\,, M_{BC}] & = - \eta_{AB} G_{CA'} + \eta_{AC} G_{BA'}\,, \\
	[G_{AA'}\, J] & = \epsilon_{A'}{}^{B'} G_{AB'}\,,
\end{align}
\vspace{-0.9cm}
\begin{align}
	\hspace{-4.3cm}[M_{AB}\,, M_{CD}] = \eta_{AC} M_{BD} - \eta_{BC} M_{AD} - \eta_{AD} M_{BC} + \eta_{BD} M_{AC}\,,
\end{align}
and
\begin{align}
	[G_{AA'}\,, P_{B'}] & = \delta_{A'B'} Z_A\,, &
	[G_{AA'}\,, G_{BB'}] & = \delta_{A'B'} Z_{AB} + \epsilon_{A'B'} \eta_{AB} S\,, \\
	[Z_A\,, M_{BC}] & = - \eta_{AB} Z_C + \eta_{AC} Z_B\,, &
	[H_A\,, Z_{BC}] & = - \eta_{AB} Z_C + \eta_{AC} Z_B\,,
\end{align}
\vspace{-0.9cm}
\begin{align}
	\hspace{-4.5cm}[M_{AB}\,, Z_{CD}] = \eta_{AC} Z_{BD} - \eta_{BC} Z_{AD} - \eta_{AD} Z_{BC} + \eta_{BD} Z_{AC}\,.
\end{align}
\end{subequations}
We will leave the explicit construction of an action for extended $p$-brane NC gravity in $p+3$ dimensions that realizes the extended $p$-brane NC algebra as the underlying symmetry algebra to future studies.

\section{Non-existence of a Non-degenerate Symmetric Invariant Bilinear Form} \label{app:form}

Let $\langle , \rangle$ be a symmetric invariant bilinear form for the ESNC algebra. We will demonstrate that $\langle , \rangle$ is necessarily degenerate because $\langle H_0 , x \rangle = 0$ for all $x$ in the algebra.

Invariance implies the associativity property $\langle [x,y] , z \rangle = \langle x , [y,z] \rangle$ for all $x,y,z$ in the ESNC algebra. We will make some judicious choices for $x$, $y$ and $z$.

Note that $\langle [ M , H_1 ] , x \rangle = - \langle H_0 , x \rangle$. On the other hand, invariance implies $\langle [ M , H_1 ] , x \rangle = \langle M , [ H_1, x] \rangle$. Therefore, $\langle H_0, x \rangle = 0$ if $[H_1, x ] = 0$, which is the case for $x = H_0, H_1 , P_2, P_3, M , G_{02}, G_{03}, J, Z_0, Z_1$, and $S$. Only three cases remain: $x = G_{12}, G_{13}$, and $Z$. For these, we have
\begin{align} \label{eq:rest}
    \langle H_0, G_{12} \rangle &= - \langle M, P_2 \rangle, &%
    \langle H_0, G_{13} \rangle &= - \langle M , P_3 \rangle, &%
    \langle H_0 , Z \rangle = - \langle M , Z_0 \rangle \,.
\end{align}
Now, $M$ commutes with $J$ and so
\begin{equation}
    0 = \langle [M , J ] , P_{A'} \rangle = \langle M, [J, P_{A'} ] \rangle = - \epsilon_{A'}{}^{B'} \langle M , P_{B'} \rangle.
\end{equation}
This implies $\langle M, P_2 \rangle = \langle M , P_3 \rangle = 0$ and thus, by eq.~\eqref{eq:rest}, that $\langle H_0, G_{12} \rangle = \langle H_0, G_{13} \rangle = 0$.

Similarly, $M$ commutes with $P_3$ and so
\begin{equation}
    0 = \langle [ M, P_3 ] , G_{03} \rangle = \langle M, [P_3, G_{03} ] \rangle = - \langle M, Z_0 \rangle.
\end{equation}
From eq.~\eqref{eq:rest}, we conclude that $\langle H_0, Z \rangle = 0$.

Using similar arguments, one can show that the kernel of any symmetric invariant bilinear form for the ESNC algebra is in fact spanned by $H_A$, $P_{A'}$ and $Z_{A}$.

\end{document}